\documentclass[10pt]{iopart}
\pdfoutput=1

\expandafter\let\csname equation*\endcsname\relax

\expandafter\let\csname endequation*\endcsname\relax

\usepackage{amsmath}

\usepackage[utf8]{inputenc}
\usepackage{multirow}

\usepackage[dvips]{graphics,graphicx}
\usepackage{url}
\usepackage{subfigure}
\usepackage{etoolbox}
\usepackage{booktabs}
\usepackage[table,xcdraw]{xcolor}
\usepackage{ulem}
\usepackage{overpic}
\usepackage{hyperref}
\usepackage{amssymb}
\usepackage{color}
\usepackage{tablefootnote}
\usepackage{nidanfloat} 

\usepackage[referable]{threeparttablex}
\usepackage{array, makecell}

\renewcommand*{\emph}{\textit}

\newtoggle{draft}

\toggletrue{draft}

\begin{document}
\title[Confinement of runaway electrons in ITER current quench]{Confinement of passing and trapped runaway electrons in the simulation of an ITER current quench}

\author{Konsta S\"arkim\"aki\textsuperscript{1}, Javier Artola\textsuperscript{1,2}, Matthias Hoelzl\textsuperscript{1}, and the JOREK team\textsuperscript{3}}

\address{\textsuperscript{1}Max Planck Institute for Plasmaphysics, 85748 Garching, Germany}
\address{\textsuperscript{2}ITER Organization, 13067 St Paul Lez Durance Cedex, France}
\address{\textsuperscript{3}See the author list of M. Hoelzl et al., `The JOREK non-linear extended MHD code and applications to large-scale instabilities and their control in magnetically confined fusion plasmas', Nucl. Fusion \textbf{61} (2021) 065001}

\ead{sarkimk1@ipp.mpg.de}
\vspace{10pt}
\begin{indented}
\item[]\today
\end{indented}

\begin{abstract}
Runaway electrons (REs) present a high-priority issue for ITER but little is known about the extent to which RE generation is affected by the stochastic field intrinsic to disrupting plasmas.
RE generation can be modelled with reduced kinetic models and there has been recent progress in involving losses due to field stochasticity, either via a loss-time parameter or radial transport coefficients which can be estimated by tracing test electrons in 3D fields.
We evaluate these terms in ITER using a recent JOREK 3D MHD simulation of plasma disruption to provide the stochastic magnetic fields where RE markers are traced with the built-in particle tracing module.
While the MHD simulation modelled only the current quench phase, the case is MHD unstable and exhibits similar relaxation as would be expected during the thermal quench.
Therefore, the RE simulations can be considered beginning right after the thermal quench but before the MHD relaxation is complete.
The plasma is found to become fully stochastic for 8~ms and the resulting transport is sufficient to overcome RE avalanche before flux surfaces are reformed.
We also study transport mechanisms for trapped REs and find those to be deconfined as well during this phase.
While the results presented here are not sufficient to assess the magnitude of the formed RE beam, we show that significant RE losses could be expected to arise due to field stochasticity.
\end{abstract}

%
\vspace{2pc}
\noindent{\it Keywords}: runaway electrons, stochastic magnetic field, vertical displacement, plasma disruption, orbit-following, ITER

%

%
%
\ioptwocol

\section{Introduction}
\label{sec:Introduction}

\begin{figure*}[b]
\centering
\begin{overpic}[width=0.99\textwidth]{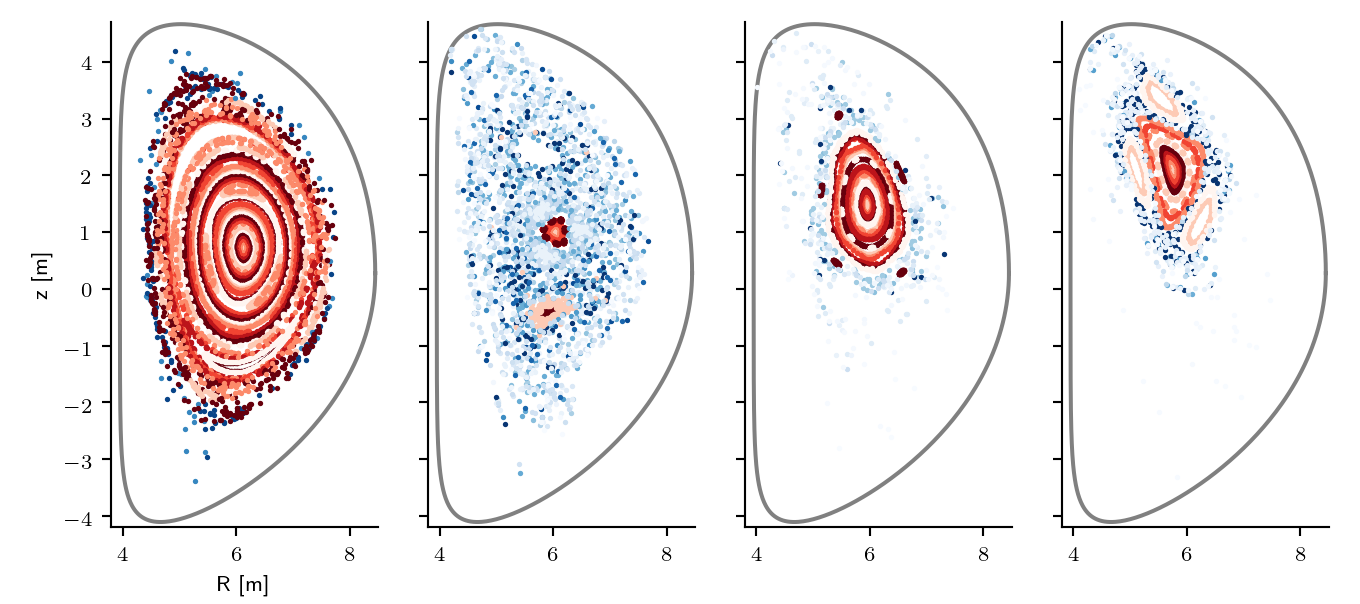}
\put(18.5,6.5){$t=10$ ms}
\put(42.0,6.5){$t=15$ ms}
\put(65.0,6.5){$t=20$ ms}
\put(88.5,6.5){$t=25$ ms}
\end{overpic}
\caption{
Magnetic field Poincaré plots at different time slices for the ITER current quench simulation being investigated in this work.
Shades of red are used to identify closed field lines.
Blue color is used for the open field lines while darker shade indicates longer connection length.
}
\label{fig: poincare}
\end{figure*}
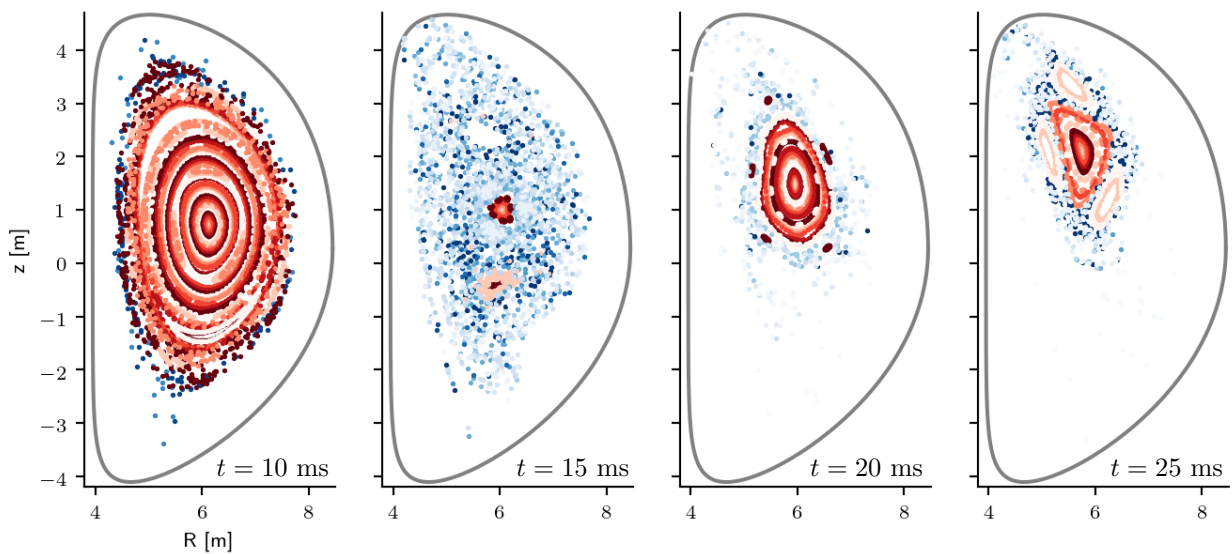

Sudden plasma cooling during tokamak disruptions causes the plasma conductivity to drop, which normally would lead to rapid dissipation of plasma current.
However, the consequent increase in electric field might be large enough that the accelerating force overcomes collisional drag experienced by fast electrons, thus creating a population of \emph{runaway electrons} (REs) that are not slowed down.

It was recognized already in the 90's that the original ``seed" population of REs born during the thermal quench quickly multiplies via an avalanching process until REs are carrying the total plasma current~\cite{Jayakumar_1993,Rosenbluth_1997}.
The current quench is therefore replaced by a current plateau, which can represent a substantial fraction of the initial plasma current being up to $\sim 10$~MA in ITER~\cite{Hender_2007,MartinSolis2017}.
RE beams are expected to be vertically unstable in ITER causing the beam to intercept the first wall~---~with deleterious melting of plasma facing components~\cite{HollmannDMS}. 

However, disruptions are MHD active events; increased RE transport can be expected due to magnetic field stochasticity during the thermal and current quenches.
The final RE beam magnitude is sensitive to the losses of RE seed population during the thermal quench~\cite{Svenningsson_2021}, and the beam is suppressed altogether if losses overcome avalanche generation during the current quench~\cite{Helander_2000, Li_2017, Martin_Solis_2021}.
Therefore, quantitative estimates for the field stochasticity induced losses are required to predict beam formation and mitigation accurately.

In this work, we investigate the confinement of runaway electrons during a current quench in ITER by tracing particles in a recently completed JOREK MHD simulation~\cite{artola_2021}.
The 3D simulation did not model thermal quench; instead, the initial conditions were assumed to be those of a mitigated disruption after the thermal quench.
Therefore for the purposes of RE modelling, the RE tracing can be considered to begin right after the thermal quench but before the accompanied MHD relaxation is complete.
During this time the initial RE beam has already been formed by the hot-tail mechanism and is being amplified via avalanche generation.

The MHD simulation (``Case 1'' in Ref.~\cite{artola_2021}) features a current quench lasting for $\sim 50$~ms, during which there is an upward vertical displacement. 
Snapshots of magnetic field structure can be found in Fig.~\ref{fig: poincare}.
The current quench begins approximately at $t=5$~ms, after the initial profiles were established in an axisymmetric simulation.
The field evolution is dominated mainly by the unstable modes with low toroidal periodicity.
Around $t=12$~ms when the dominant $n=1,2,3$ modes peak, the plasma becomes almost fully stochastic. 
The stochastic phase lasts for $\sim 8$~ms after which flux surfaces in the core reform when the core safety factor exceeds the $q_s=2$ resonance.
The simulation did not consider REs, which is justified if the RE current is negligible before the stochastic phase begins and the field stochasticity suppresses further generation.

Therefore, the aim of this study is to assess the losses of REs during the stochastic phase and whether there is potential for some REs to survive this phase to form a beam when flux surfaces have reformed.
The assessment is done in three parts:
\begin{itemize}
    \item 
    Particles in this study are traced with the JOREK built-in orbit-following module. 
    For more realistic RE simulations, the code is retrofitted with operators for Coulomb collisions and synchrotron losses, which we introduce and verify (Section~\ref{sec:orbitfollowing}).
    \item
    Orbit-following simulations are done to assess RE confinement during the stochastic phase.
    Losses are quantified by computing transport coefficients, and the results are used to determine whether the losses are sufficient to overcome avalanche generation (Section~\ref{sec:passing}).
    \item
    The evaluation of transport coefficients is limited to only passing REs but poloidally trapped REs can be created during the thermal quench and via avalanche.
    Trapped particles are not transported along the stochastic field lines so there is a possibility of them surviving the stochastic phase.
    The confinement of trapped REs in general is relatively unexplored topic which we study separately (Section~\ref{sec:trapped}).
\end{itemize}

\section{The orbit-following model}
\label{sec:orbitfollowing}


The JOREK particle tracer has the capability to trace relativistic particles in time-evolving fields, calculated by the non-linear 3D MHD code JOREK~\cite{hoelzl2021jorek}, either by solving the full gyromotion, or the corresponding guiding-center motion~\cite{Sommariva_2017}.
The gyromotion is solved with the Volume-preserving algorithm~\cite{zhang2015volume}, and the guiding-center equations of motion~\cite{Tao_2007} with RK4.
While the latter scheme does not exactly preserve energy, the time step was chosen to be small enough that this error remained below 0.1~\% in the simulations performed in this work.

\subsection{Collision operator}

In earlier work, the collisional drag force acting on REs was implemented to the JOREK particle tracer~\cite{Sommariva_2018}. 
Here we have replaced it with a full collision operator including both collisional drag and diffusion in energy, and also the pitch scattering term.
The new collision operator has its basis on the relativistic Braams~-~Karney collision operator~\cite{Braams_1987}, which is simplified for orbit-following purposes by assuming a test particle moving through an isotropic plasma in thermal equilibrium~\cite{Pike_2014}.
Numerical implementation follows closely Ref.~\cite{Sarkimaki_2018}, and here we only review the main aspects.
The model introduced here does not account for the screening effect relevant when partially ionized impurities are present in the plasma~\cite{Hesslow_2018a}, and its implementation is left for future work.

Written in Fokker-Planck form, the collision operator for test particle species $a$ reads
\begin{equation}
\label{eq: fokker planck particle}
\left(\frac{\partial f_a}{\partial t}\right)_\mathrm{coll} = -\frac{\partial}{\partial \mathbf{p}}\cdot(\mathbf{K}_{ab} f_a)
+ \frac{\partial}{\partial \mathbf{p}}\frac{\partial}{\partial \mathbf{p}}:(\mathbf{D}_{ab} f_a),
\end{equation}
where $f$ is the distribution function, $t$ the time, and $\mathbf{p}$ the momentum normalized to particle rest mass times speed of light $m_a c$.
The Fokker-Planck drift, $\mathbf{K}_{ab}$, and diffusion, $\mathbf{D}_{ab}$, coefficients in their general form are functions of the background species (labeled $b$) distribution function $f_b$.
When $f_b$ is isotropic, the drift coefficient can be written as $\mathbf{K}_{ab} = K_{ab}\mathbf{\hat{p}}$, and the diffusion coefficient can be separated into parallel and perpendicular terms as $\mathbf{D}_{ab} = D_{\parallel,ab}\mathbf{\hat{p}}\mathbf{\hat{p}} + D_{\perp,ab}(\mathbf{I} - \mathbf{\hat{p}}\mathbf{\hat{p}})$, where $\mathbf{I}$ is the identity matrix.

Further assuming Maxwell-J\"uttner distribution for the background species, the coefficients become~\cite{Pike_2014}
\begin{align}
K_{ab}           &= -\Gamma_{ab}\frac{1}{p^3}\left(\frac{\mu_0}{\gamma} + \frac{m_a}{m_b}\mu_1 \right),\\
D_{\parallel,ab} &= \Gamma_{ab}\frac{\Theta_b\gamma}{p^3}\mu_1,\\
D_{\perp,ab}     &= \Gamma_{ab}\frac{1}{2\gamma p^3}\left(p^2(\mu_0 + \gamma\Theta_b\mu_2) - \Theta_b\mu_1\right),
\end{align}
where $\Gamma_{ab} \equiv $ $n_b q_a^2 q_b^2 \ln\Lambda_{ab} / 4\pi\epsilon_0^2 m_a^2c^3$, $n_b$ is the number density, $q$ the particle charge, $\ln\Lambda$ the Coulomb logarithm, $\gamma = \sqrt{1+p^2}$, and $\mu_0(p,\Theta_b)$, $\mu_1(p,\Theta_b)$, and $\mu_2(p,\Theta_b)$ are functions of $p$ and normalized temperature $\Theta_b\equiv T_b/m_bc^2$.
Explicit forms for the special functions and their numerical evaluation can be found in Ref.~\cite{Sarkimaki_2018}.

The Langevin equation corresponding to the collision operator, Eq.~\eqref{eq: fokker planck particle}, is
\begin{multline}
d\mathbf{p} = K \mathbf{\hat{p}}dt \\+ \left[\sqrt{2D_{\parallel}}\mathbf{\hat{p}}\mathbf{\hat{p}}\right. 
+ \left.\sqrt{2D_{\perp}}(\mathbf{I} - \mathbf{\hat{p}}\mathbf{\hat{p}}) \right]\cdot d\mathbf{W},
\end{multline}
where the coefficients are summed over all background species: $K \equiv \sum_b K_{ab}$, $D_\parallel \equiv \sum_b D_{\parallel,ab}$, and $D_\perp \equiv \sum_b D_{\perp,ab}$.
Thus the change in test particle momentum due to collisions is given by a stochastic differential equation where $\mathbf{W}(t)$ represents a three-dimensional Wiener process with zero mean and variance $t$.
The Langevin equation is discretized with the Euler-Maruyama method by substituting $dt\rightarrow \Delta t$ and $d\mathbf{W}\rightarrow \sqrt{\Delta t}\mathbf{X}$, where each element of $\mathbf{X}$ has the equal probability of being $-1$ or $+1$ for each realization.
At each time step in the simulation loop, markers are advanced by first solving the gyromotion due to the background field, and then separately evaluating collisions.

The guiding-center equations of motion are obtained by applying Lie transform perturbation methods to the Hamiltonian of a charge particle~\cite{Tao_2007}.
When the same transform is applied to the particle Fokker-Planck equation, and the result is gyro-averaged to remove the fast time scale, the guiding-center Fokker-Planck equation is obtained~\cite{Brizard_2004,Hirvijoki_2013}.
The corresponding guiding-center Langevin equations are
\begin{align}
dp &= \mathcal{K} dt + \sqrt{2D_{\parallel}}dW_p,\\
\label{eq: coll pitch}
d\xi &= -\nu\xi + \sqrt{(1-\xi^2)\nu}dW_\xi
\end{align}
where $\xi$ is guiding center pitch, $W_p$ and $W_\xi$ are independent Wiener processes,
\begin{equation}
\mathcal{K} \equiv -\Gamma_{ab} \frac{m_a}{p^2m_b}\mu_1 + \frac{dD_\parallel}{dp} + \frac{2D_\parallel}{p},
\end{equation}
is the collisional drag, and
\begin{equation}
\label{eq: pitch scattering}
\nu \equiv \frac{2D_{\perp}}{p^2},
\end{equation}
is the pitch collision frequency.
For numerical implementation, these equations are also discretized with the Euler-Maruyama method as was done in the particle picture.

Including collisions in the guiding-center picture requires the coordinate transformation ($p_\parallel$, $\mu$) $\rightarrow$ ($p$, $\xi$), and its inverse, because the Hamiltonian motion is solved with parallel momentum, $p_\parallel$, and magnetic moment, $\mu$, as the momentum space coordinates.
It is possible to solve collisions in ($p_\parallel$, $\mu$) basis, but this would result in a non-diagonal diffusion coefficient.
This has the drawback that one would not be able to treat pitch scattering and energy collisions separately, which is useful for simulation analysis, and it would also complicate implementation of adaptive time-stepping~\cite{Sarkimaki_2018} if that is sought in future development.

Finally, we have omitted the Langevin equation corresponding to the guiding-center spatial coordinate whose role is to account for the classical transport.
Since neoclassical transport is already included in our model, via Eq.~\eqref{eq: coll pitch}, and it dominates over the classical transport in tokamaks, we can safely omit the spatial operator.
Because the collision operator does not affect the guiding-center position, we can avoid one computationally expensive magnetic field evaluation which would be needed otherwise for the ($p$, $\xi$) $\rightarrow$ ($p_\parallel$, $\mu$) conversion performed after the collisions have been evaluated.

\subsection{Radiation reaction force}

In addition to collisions, another energy loss mechanism relevant for REs is synchrotron emission.
When radiation is emitted, the radiation reaction force is exerted to the particle which, in addition to causing energy loss, could have an effect on particle transport.

The model for the radiation reaction force for particle and guiding center dynamics is taken from Ref.~\cite{Hirvijoki_2015}.
For (full orbit) particles this reads
\begin{equation}
\label{eq: rad force particle}
d\mathbf{p} = -\tau_r^{-1}\left(\mathbf{p}_\perp + p_\perp^2\mathbf{p} \right)dt,
\end{equation}
and for guiding center
\begin{align}
\label{eq: rad react ppar}
dp_\parallel &= \tau_r^{-1}p_\parallel\frac{2\mu B}{mc^2}dt,\\
\label{eq: rad react mu}
d\mu &= -\tau_r^{-1}\mu\left(2 + \frac{4\mu B}{mc^2}\right)dt,
\end{align}
where $B$ is the magnetic field magnitude and
\begin{equation}
\tau_r \equiv \frac{6\pi\epsilon_0\gamma(mc)^3}{q^4B^2},
\end{equation}
is the characteristic time for the radiation reaction force.

The guiding-center radiation reaction force is readily compatible with RK4 used to solve the Hamiltonian motion.
However, this is not the case for the particle motion as the Volume-preserving algorithm is used.
Because the Hamiltonian motion is dominant and the effect of the radiation reaction force is comparable to the Coulomb collisions, we resolve to use the Euler method for solving Eq.~\eqref{eq: rad force particle}.
For consistency the guiding-center radiation reaction force is also solved with the Euler method in this work.

\begin{figure}[t]
\centering
\begin{overpic}[width=0.49\textwidth]{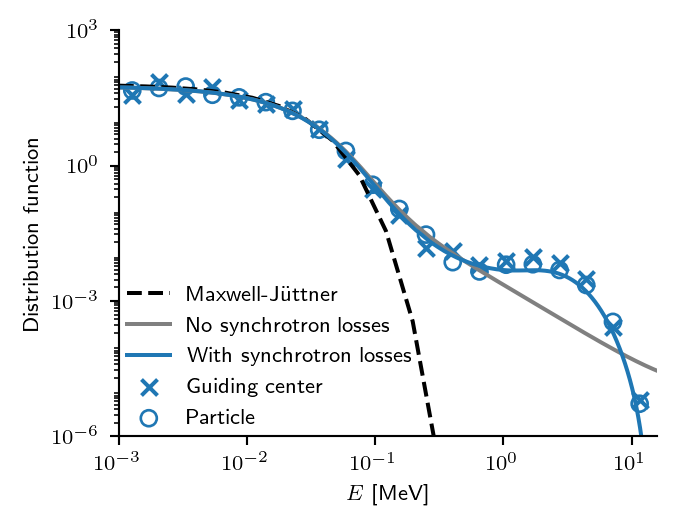}
\end{overpic}
\caption{
Verification of the bump-on-tail formation.
The solid lines correspond to the DREAM simulations: synchrotron losses included (blue) and without (grey) shown for reference.
The markers correspond to the JOREK particle tracer results when using the particle (circles) or guiding center (crosses) mode.
The dashed line is the initial Maxwell-J\"uttner distribution used in both DREAM and JOREK particle simulations.
}
\label{fig: bumpontail}
\end{figure}

\subsection{Verification}

We verify the new developments by using the particle tracer to reproduce the so-called \emph{bump-on-tail distribution}~\cite{Decker_2016} which arises when both collisional dynamics and synchrotron losses are accounted for.
Collisional diffusion near the critical energy (defined as the value where electric field acceleration first overcomes friction) creates constantly new runaway electrons when the electric field is above the critical value, $E_\mathrm{crit}=ne^3\ln\Lambda / 4\pi\epsilon_0^2mc^2$, required for RE generation.
Instead of being accelerated indefinitely, REs at high energy experience significant synchrotron losses that are enhanced by collisional scattering.
Therefore REs accumulate between the critical energy and the maximum energy set by the radiation reaction force, eventually forming a bump on the tail of the electron energy distribution.

For this test, 3000 electron markers sampled from the Maxwell-J\"uttner distribution are initialized in a hydrogen plasma, with fixed $T=16.2$~keV and $n=10^{19}$~m$^{-3}$, and traced for 10~s.
The magnetic field is cylindrical with $B=4$~T and the electric field is constant at $E_\parallel=3\times E_\mathrm{crit}\approx 0.03$ V/m.
The test case is adapted from Ref.~\cite{Decker_2016}, where the bump-on-tail distribution was studied with the kinetic code CODE~\cite{Landreman_2014}.

Since there is no analytical formula for the bump-on-tail distribution, the simulation result is verified by comparing it to the kinetic code DREAM~\cite{Hoppe_2021}, which has been thoroughly benchmarked to CODE and can be viewed as CODE's successor. 
Figure~\ref{fig: bumpontail} shows the comparison and we find there is a good agreement between DREAM and the JOREK particle tracer in both modes.

\section{Runway-electron losses during the stochastic phase}
\label{sec:passing}

An open problem in RE studies is to couple losses due to the stochastic field to RE generation~\cite{Svensson_2021,Tinguely_2021}, and RE beam evolution to MHD equations~\cite{Harvey_2019,liu2020structure,Bandaru_2021}.
Self consistent treatment of the whole RE dynamics would therefore require coupling between the orbit-following method, kinetic tools, and MHD codes~\cite{Hirvijokifluid_2018}.
This is beyond the scope of this work; instead, here we evaluate drift and diffusion coefficients associated with the radial transport due to stochastic field lines which in future work could be used in reduced kinetic codes such as DREAM.
Here we use these coefficients to estimate the loss time which allows to assess whether the transport during the current quench phase is sufficient to overcome avalanche generation~\cite{Martin_Solis_2021}.

\subsection{Transport coefficient evaluation}

The transport coefficients are calculated by tracing electrons for several orbit circulation times. 
To only account for the transport due to magnetic field perturbations, collisions and radiation reaction force are disabled and electric field is artificially set to zero (their effect on transport is studied separately).
Initially all electron markers have same radial coordinate $r$, momentum $p$, and pitch $\xi$, but these, as well as the time instance $t$ when markers are initialized, are varied between the simulations to scan the parameter space ($r$, $p$, $\xi$, $t$).
Toroidally the markers are distributed uniformly.
The radial coordinate is defined as distance to the magnetic axis measured at the outer mid-plane.

From the results of each simulation, we compute advection, $K_s$, and diffusion, $D_s$, coefficients which have been shown to model RE transport in a stochastic field with good accuracy~\cite{Sarkimaki_2016}.
In other words, we assume that the marker radial position, $r_i$, obeys the stochastic differential equation
\begin{equation}
\label{eq: random walker}
dr_i = K_s dt + \sqrt{2D_s} dW.
\end{equation}
An advection coefficient is required as pure diffusion, i.e. Rechester-Rosenbluth model~\cite{Rechester_1978}, is insufficient to capture the transport accurately~\cite{Papp_2011PPCF}.
Coefficients are evaluated separately for markers that remain confined for the duration of the simulation and for those that are lost. 
A mean value is used to represent transport at the given phase space location as
\begin{equation}
A \equiv \frac{1}{N_\mathrm{tot}}\sum_i^{N_\mathrm{conf}} A_i 
+ \left(1-\frac{N_\mathrm{conf}}{N_\mathrm{tot}}\right)A_\mathrm{lost},
\end{equation}
where $A$ is either $K_s$ or $D_s$, and $N_\mathrm{conf}$ and $N_\mathrm{tot}$ are number of confined and total number of markers, respectively, and $A_i$ and $A_\mathrm{lost}$ are defined below.

For all \emph{lost} markers we compute common values for the transport coefficients based on the distribution of their \emph{loss times} $t_i$, i.e., the time it took for marker $i$ to become lost after it was initialized.
This distribution is analogous to the so-called \emph{first passage time distribution}, which is a distribution of times when markers launched from the same position first pass a fixed position (e.g. the separatrix).
For a random walker obeying Eq.~\eqref{eq: random walker}, this distribution is given by
\begin{equation}
\label{eq: first passage time}
T(t) = \frac{\Delta r}{\sqrt{\pi D_s t^3}}\exp\left(-\frac{\left(\Delta r - K_st\right)^2}{D_st}\right),
\end{equation}
where $\Delta r$ is the distance from the initial marker position to the fixed position, which in our case is the distance to the separatrix at the outer mid-plane.
Now we can use the statistical properties of the first passage time distribution to evaluate the advection and diffusion coefficients:
\begin{align}
K_\mathrm{lost} &= \frac{\Delta r}{ \mathrm{Mean}(t_i) },\\
D_\mathrm{lost} &= \frac{(\Delta r)^2}{2}\frac{ \mathrm{Var}(t_i)}{[\mathrm{Mean}(t_i)]^3 }.
\end{align}

For \emph{confined} markers, the coefficients are evaluated by recording the radial coordinates where markers pass the outer mid-plane.
Coefficients are then simply computed as
\begin{align}
K_i &= \frac{\mathrm{Mean}(r_{i,j} - r_{i,j-1})}{\Delta t}, \\
D_i &= \frac{\mathrm{Var}(r_{i,j} - r_{i,j-1})}{2\Delta t},
\end{align}
where $r_{i,j}$ is the $j$'th passing of marker $i$, and $\Delta t$ is the average time between subsequent passings.
However, this computation may show artificial diffusion if a marker is confined within a remnant magnetic island: the marker can jump from the inner boundary to the outer, and vice versa, between subsequent outer mid-plane passings, yielding non-zero $\mathrm{Var}(r_{i,j}-r_{i,j-1})$ even though no actual transport is present.
This artificial noise in the diffusion term can be somewhat reduced by replacing $r_{i,j}$ with its mean value between $n$ subsequent passings.

\subsection{Results and discussion}

\begin{figure}[t]
\centering
\begin{overpic}[width=0.49\textwidth]{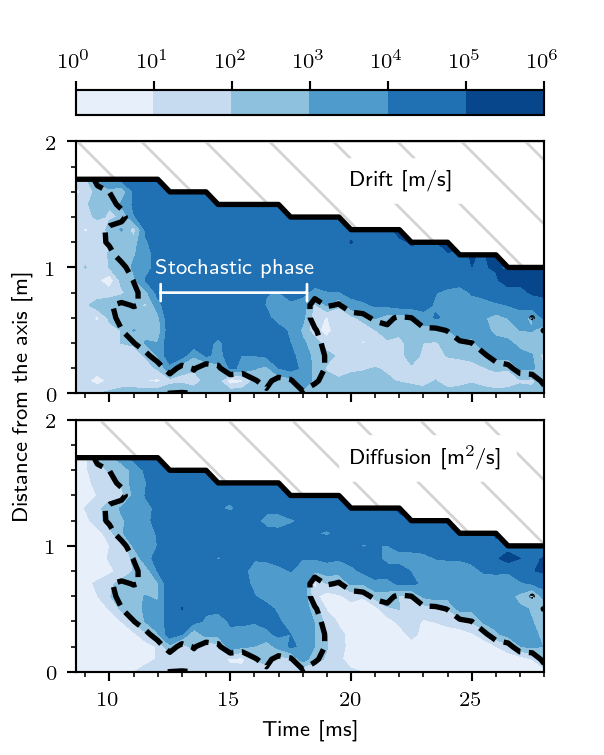}
\end{overpic}
\caption{
Advection and diffusion coefficients calculated from orbit-following simulations as a function of time and energy.
The dashed black contour shows the region where the loss time $\tau_s < 1$~ms according to the coefficient-based estimation.
Hatched region is beyond the last closed flux-surface; the coarse step-like structure appears at the edge because the computational grid had a radial bin size of approx. 10 cm.
The results shown here are for electrons with $E_\mathrm{kin}=200$~keV and $p_\parallel/p=0.9$.
Scans were performed for $E_\mathrm{kin}=$ 3~keV--~50~MeV and $p_\parallel/p =$ 0.8~--~1.0 and the region where $\tau_s < 1$~ms was approximately the same in each case.
}
\label{fig: coefficients}
\end{figure}

The transport coefficients evaluated for the ITER case are shown in Fig.~\ref{fig: coefficients} as a function of time and radius.
Initially there is no transport as the flux surfaces are intact, but between 10~--~12~ms (when the mode energies peak, see Fig.~4 in~\cite{artola_2021}) the stochastic field region expands until it penetrates almost the whole plasma.
This is the beginning of the stochastic phase; also here the confinement volume begins to shrink due to plasma moving vertically into the wall.
Around $t=18$~ms the stochastic phase ends as flux surfaces reform in the core and only the edge from $r>0.8$~m remains stochastic.
The width of the stochastic layer keeps roughly constant as the confinement volume shrinks further. 
At the end of the simulation, the whole plasma is again stochastic except for the remnant magnetic islands appearing at $t=23$~ms which reduce transport in that region.

For losses to overcome the avalanche process, the loss time needs to be smaller than 1~--~10~ms~\cite{Martin_Solis_2021}.
The loss time, $\tau_s$, is defined from the relation $N=N_0\exp(-t/\tau_s)$, where the number of REs in the plasma, $N$, is assumed to decay exponentially due to stochastic losses.
When calculating the transport coefficients, we assumed the transport to be a combination of advection and diffusion, in which case the rate of losses is not constant in time but given by the first passage time distribution, Eq.~\eqref{eq: first passage time}.
The distribution is peaked and we can use the time instance at the peak location to estimate the loss time from the transport coefficients.
Finding the roots of Eq.~\eqref{eq: first passage time}, we obtain an estimate for the loss time:
\begin{equation}
\label{eq: loss time}
\tau_s = \frac{\sqrt{16\Delta r K_s^2 + 9 D_s^2} - 3D_s}{4K_s^2}.
\end{equation}
We set $\Delta r = 1$~m to visualize the region where the conservative limit $\tau_s < 1$~ms is met  (Fig.~\ref{fig: coefficients}).
In this region the stochastic field transport mitigates avalanche and this covers most of the plasma volume during the stochastic phase.

The only region where the condition $\tau_s < 1$~ms is not met is near the axis where transport is several orders of magnitude lower.
It is possible that even this low transport is only an artefact from the transport coefficient evaluation, and in reality the particles are trapped in the core island(s) that is present at least at $t=15$~ms according to the Poincaré plot (recall Fig.~\ref{fig: poincare}).

To make an accurate assessment of RE confinement near the axis, electron markers are initialized toroidally uniformly near the axis at $t=12$~ms and traced, this time with collisions and synchrotron losses included, for 8~ms until the stochastic phase ends.
We choose the electron initial energy to be somewhat above the critical momentum,
\begin{equation}
p_\mathrm{crit} = \frac{1}{\sqrt{E_\parallel/E_\mathrm{crit}-1}}.
\end{equation}
On axis we have $E_\parallel\approx 60$~V/m and $E_\mathrm{crit} \approx 0.8$~V/m, hence $p_\mathrm{crit}\approx 0.1$ corresponding to 3~keV, and we choose 60~keV as the initial energy.
While this choice is arbitrary, the exact value is of little relevance because the electrons are quickly accelerated to the MeV range due to the strong parallel electric field.

\begin{figure*}[t]
\centering
\begin{overpic}[width=0.99\textwidth]{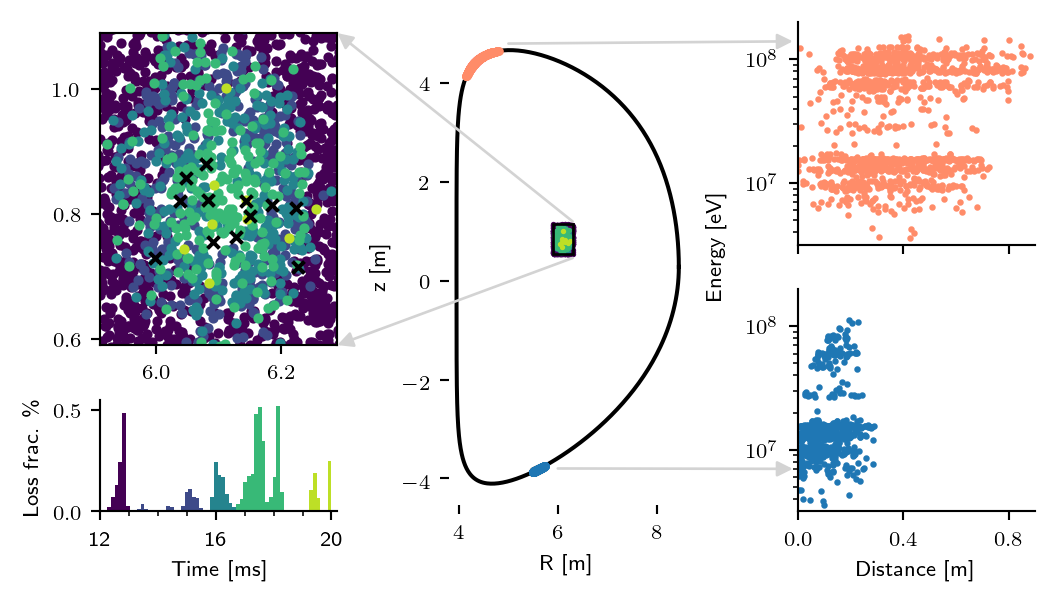}
\put( 5,55){a)}
\put(40,55){c)}
\put( 5,21){b)}
\put(76,55){d)}
\put(76,30){e)}
\end{overpic}
\caption{
Confinement of REs that are located close to the axis when the stochastic phase begins.
(a) Scatter plot showing markers' initial $(R,\;z)$~-~coordinates.
Color corresponds to the time when that marker was lost, shown in (b), and markers that remained confined as shown with black crosses.
In (b) each bar shows what fraction of particles remaining at that time instant were lost.
The poloidal cross section (c) illustrates the size of the core region and also shows the markers' final coordinates (red and blue regions) on the edge of the computational domain, i.e. wall (black line).
Markers colored in red were lost near the location on the wall where plasma drifts, and ones colored in blue were lost to the divertor.
For both populations the wetted region as a function of final energy is illustrated in (d) and (e), where the distance is measured from the left-most impact point along the wall.
}
\label{fig: beam}
\end{figure*}
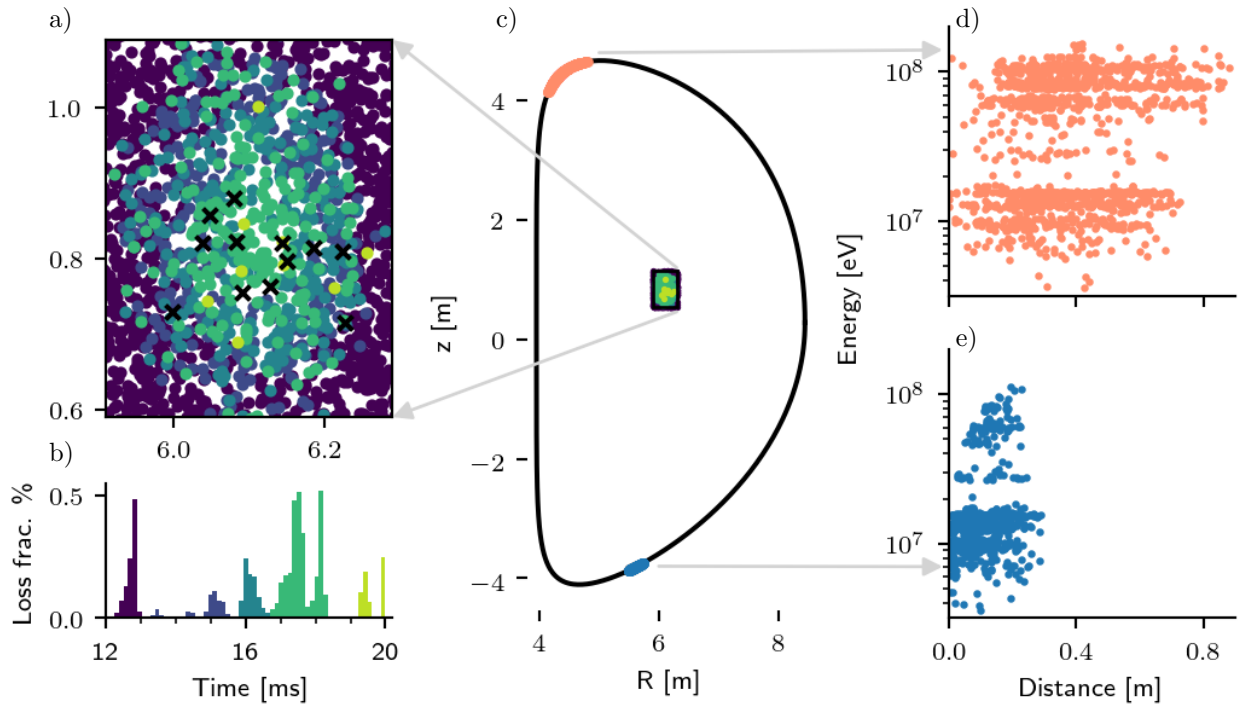

The results are gathered in Fig.~\ref{fig: beam}.
Markers that are lost within 1~ms (dark blue in (a) and (b)) allow us to deduce the extent of the region of low transport.
The region is found to be approximately 30~cm~$\times$~50~cm in size.
However, even within this region only a small fraction of the particles survive to the end of the stochastic phase.

The number of simulated markers was increased until approximately 1000 markers remained in the low-transport region at $t=13$~ms after the initial losses.
Instead of ``leaking'' at a constant rate, these markers were found to be lost in short bursts at $t=15$~ms (shown in blue), $t=16$~ms (teal), $t=17$~ms (turquoise) and $t=19$~ms (yellow).
Except for the last pulse, each pulse ejects particles further in from the plasma thus eliminating any RE beam that could have formed in that region.

The markers that survive to the end of the stochastic phase (marked with crosses) do not show any coherent structure in their initial positions, indicating no island exists that would remain intact for the whole duration.
All in all, less than 2~\% of the 1000 markers survive to the end of the stochastic phase.
The actual fraction of REs that survive is likely significantly lower than this since all REs outside the axis region are lost.
However, precise estimate for the survival fraction is not possible without knowing the radial distribution of REs at the beginning of the stochastic phase.

The final state of the lost markers at the edge of the computational domain is also shown in Fig.~\ref{fig: beam}.
All markers are lost either to the location on the upper-left corner of the wall into which the plasma is drifting (shown in red) or to the divertor leg at the low-field side (blue).
The toroidal distribution of the markers on the wall is uniform, and poloidally the wetted region is almost 0.8~m in size.
On the divertor the wetted region is approximately 0.3~m.
The marker deposition on the wall seems to shift towards the low-field side with increasing particle energy due to the final orbit width effects.
However, this shift could also be due to the vertical motion of the plasma considering that marker energy strongly correlates with time; all markers had same initial energy and the plasma parameters are roughly constant in the region where markers were initialized.

\section{Transport of poloidally trapped runaway electrons}
\label{sec:trapped}

REs are usually assumed to be on passing orbits as the electric field does not result in net acceleration for poloidally trapped particles.
There can, however, exist superthermal electrons with $p > p_\mathrm{crit}$ that are poloidally trapped.
In fact, such REs are generated in significant numbers during the avalanche~\cite{Embreus_2018,Nilsson_2015,Nilsson_2015a}: thermal electrons lifted to the RE regime via knock-on collisions can have large $p_\perp$ causing them to become poloidally trapped when they are born off-axis.
Also during the hot-tail generation the initial fast electron population is isotropic.

The issue with poloidally trapped REs is that they do not travel for a long distance along stochastic field lines.
As such, a possibility exists that they remain confined even if the plasma becomes momentarily stochastic and passing RE inventory is lost.
When flux surfaces are reformed, some of the surviving trapped REs could turn to passing orbits via collisional scattering or Ware pinch~\cite{Nilsson_2015} and provide a new RE seed.

The stochastic phase lasts for 8~ms in the ITER simulation studied here.
It is therefore unlikely that there would be poloidally trapped REs with high enough energy to survive this phase without thermalizing.
Admittedly, a more relevant case to study trapped RE confinement would be at the end of the thermal quench in cases where the stochastic phase is brief.
Nevertheless, the conditions here at the onset of the stochastic phase ($n \approx 10^{21}$~m$^{-3}$, $E_\parallel \approx 60$~V/m, and $\delta B / B \approx 10^{-2}$) are typical to those seen at the end of thermal quench, and performing an exploratory study on trapped RE confinement could provide results that are generalizable to other cases.

\subsection{Loss mechanisms}

\begin{figure*}[t]
\centering
\begin{overpic}[width=0.99\textwidth]{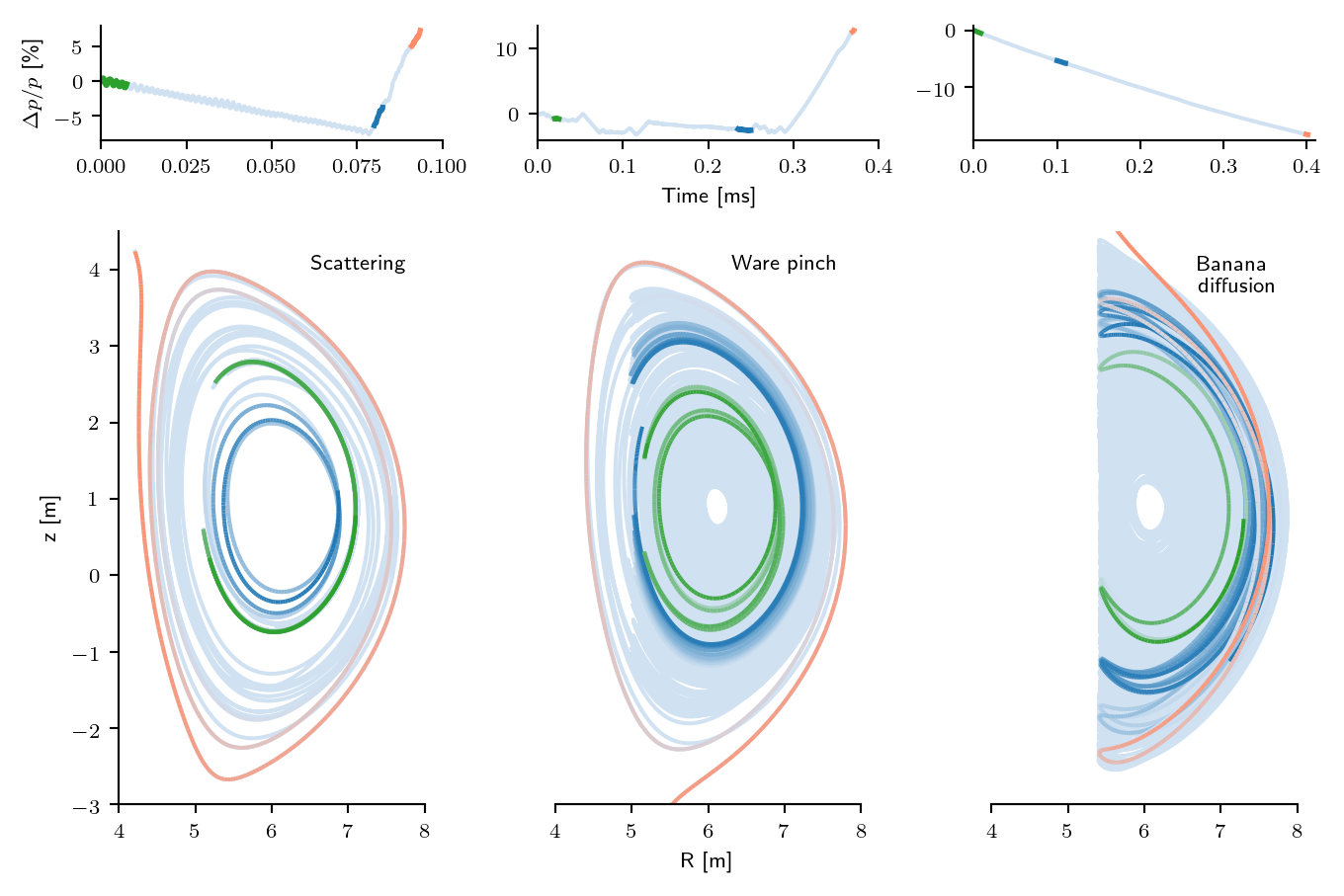}
\end{overpic}
\caption{
Orbit samples illustrating the mechanisms by which trapped electrons can be lost.
Top row shows the change in electron momentum as a function of time, and the plots on the bottom show the orbit projection on $Rz$-plane.
The complete orbit is shown in light blue and parts of the orbit are highlighted (same parts in both $p(t)$ and $(R,\;z)$ plots) with different color.
}
\label{fig: orbits}
\end{figure*}

In orbit-following simulations, we observed three dominant mechanisms that led to transport and losses of trapped REs.
For purposes of illustration, three sample RE markers were traced at the beginning of the stochastic phase and their trajectories~---~from birth to lost~---~ are shown in Fig.~\ref{fig: orbits}.
Particles had initially different energies which affected the dynamics in each case:
\begin{enumerate}
    \item 
    \emph{Collisional scattering (Fig.~\ref{fig: orbits} left column).}
    At first, a low-energy RE ($E_\mathrm{kin} = 10$~keV) executes its banana orbit for numerous times without noticeable change in trajectory (green).
    The RE is slowing down by the collisional drag, which increases its pitch collision frequency until the electron abruptly becomes passing (blue).
    This leads to a sharp increase in energy, the pitch gets reduced, and the particle does not become poloidally trapped anymore as it is soon lost by the field stochasticity (red).
    \item
    \emph{Ware pinch (Fig.~\ref{fig: orbits} middle column).}
    A mid-energy RE ($E_\mathrm{kin} = 100$~keV) is initially marginally-trapped poloidally, where it alternates between passing and trapped orbits (green).
    The electric field plays a prominent role in this mechanism by accelerating and deaccelerating the particle depending on the direction the particle is travelling along the field.
    This alternating acceleration causes the lower banana turning point to be the one that is always closer to the high-field side, which leads to inward transport when the particle is poloidally trapped. This effect is known as the Ware pinch~\cite{Nilsson_2015}.
    As the particle moves inwards, the parallel momentum required to switch from trapped to passing orbit decreases and the particle becomes passing (blue).
    Depending on the sign of $p_\parallel$ when the particle becomes passing, the particle can either gain $p_\parallel$ on the passing orbit, allowing it to travel further outwards before it becomes trapped, or lose it, causing it to become trapped sooner.
    The net effect is outward transport of REs though many trapped-passing cycles might be required until it is lost (red) as in the sample case shown here.
    \item
    \emph{Collisionless banana diffusion (Fig.~\ref{fig: orbits} right column).}
    A high energy RE ($E_\mathrm{kin} = 10$~MeV) has a wide orbit (green).
    This makes it susceptible to toroidal variation in poloidal field strength, which causes displacement of the banana turning point (blue).
    The process is analogous to ripple diffusion, where the displacement is due to toroidal variation of toroidal field strength~\cite{Goldston_1981}, and likewise this leads to decorrelation and transport~\cite{Sarkimaki_2018a}.
    This is the only loss mechanism where the particle does not become passing first before it is lost (red).
    Even though the particle energy is decreasing due to synchrotron losses, this does not affect the transport significantly.
\end{enumerate}

A fourth mechanism is possible if the electric field is sufficiently high to accelerate trapped particles to passing orbits within the time it takes to complete the half-orbit.
This mechanism would be more prominent for low-energy electrons, but those also experience significant pitch scattering that can interrupt this process.

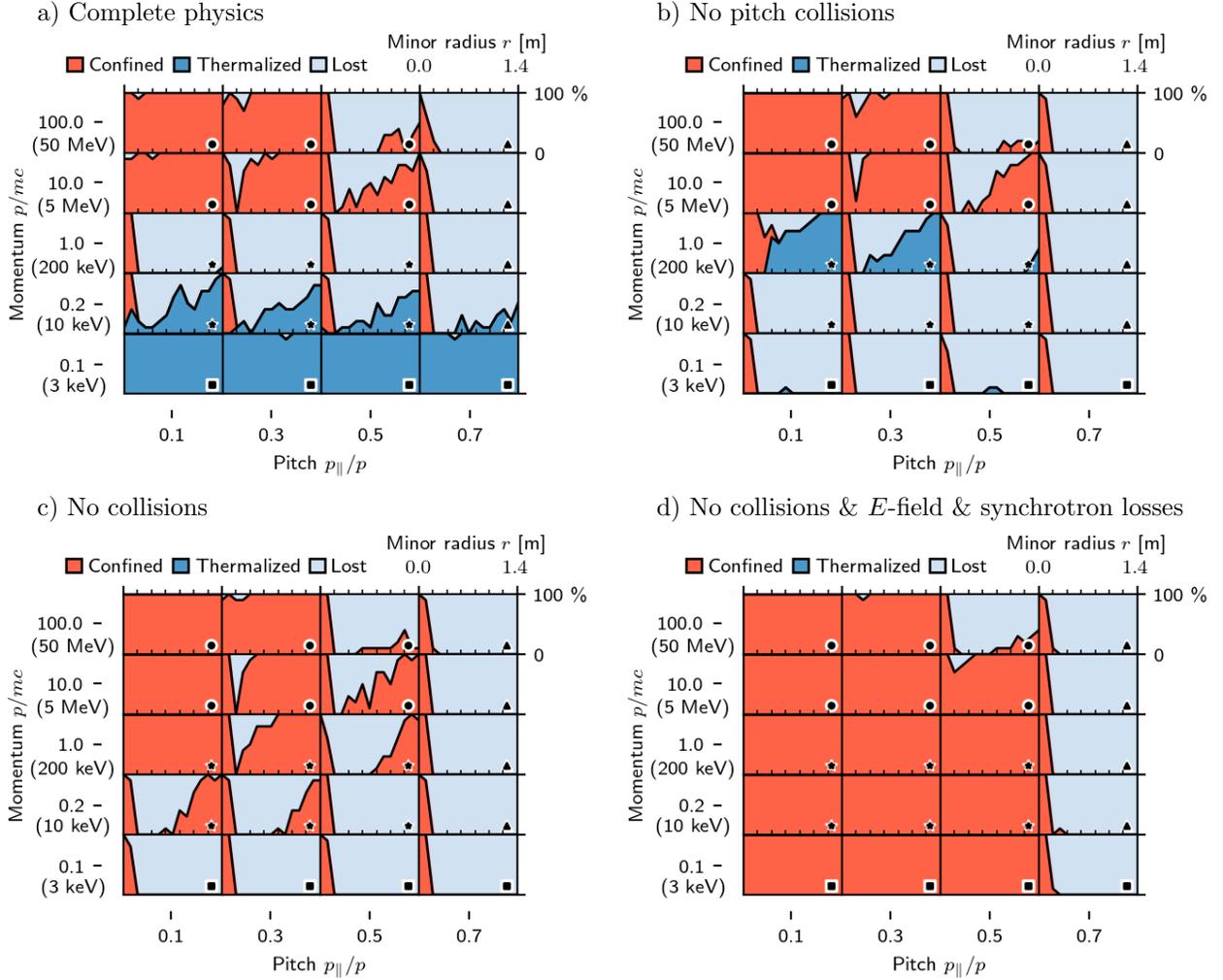
\begin{figure*}[t]
\centering
\begin{overpic}[width=0.99\textwidth]{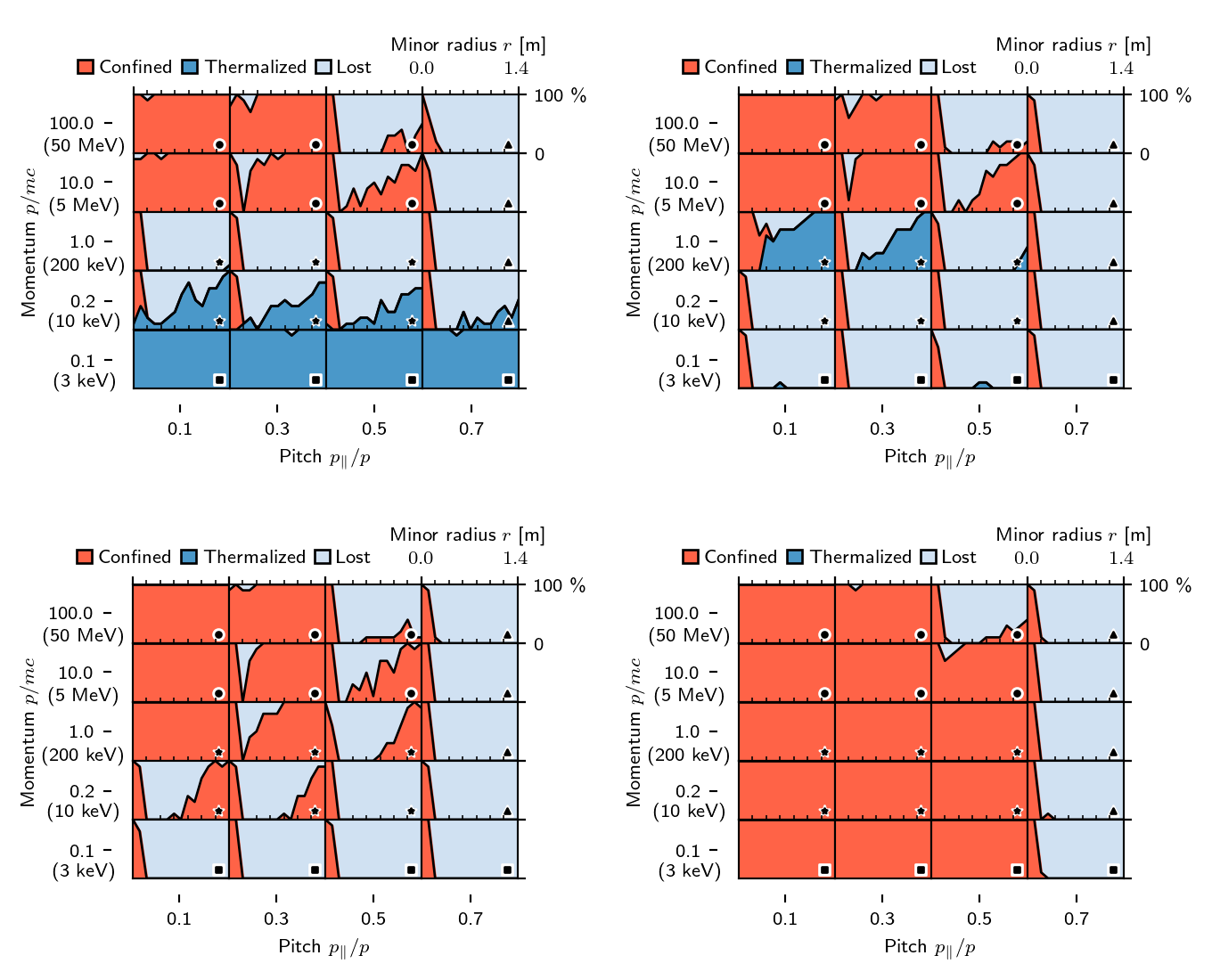}
\put(4,79){a) Complete physics}
\put(54,79){b) No pitch collisions }
\put(4,39){c) No collisions}
\put(54,39){d) No collisions \& $E$-field \& synchrotron losses}
\end{overpic}
\caption{
End state of electrons which were traced between 12.5~--~13.5 ms. 
Colours show the end state, i.e. whether an electron thermalized, was lost, or remained superthermal and confined, as a function of initial radial coordinate ranging from $r=0$ to $r=1.4$ m.
In other words, the vertical span of each coloured region is the probability of a particle from that radial position reaching one of the aforementioned outcomes.
Each panel shows the result for an electron population with different initial momentum and pitch.
The physics that were included in the simulations were varied between the subplots: (a) complete physics were included, (b) pitch collision operator was disabled, (c) the whole collision operator was disabled, and (d) the collision operator, the electric field and also the radiation reaction force were disabled.
The symbols in the panels are used to refer to different regions in phase space in the text, and they are same in each panel between the subplots.
}
\label{fig: endstate}
\end{figure*}

The characteristic time-scales for the mechanisms introduced here are collected in Table.~\ref{tab: timescales}, along with their numerical estimates.
The time-scale for the collisional scattering, $\tau_c$, is estimated from the pitch collision frequency.
A particle that initially has $p_\parallel=0$ and $p_\perp = p$ is accelerated by the electric field to $p_\parallel = p_\perp$ in time $\tau_e$, which we simply estimate from the Lorentz force as $p = (q \mathbf{E} / mc)\tau_e$.
The time it takes for a trapped particle to become passing due to the Ware pinch, $\tau_w$, was estimated in Ref.~\cite{Nilsson_2015}.
For the collisionless banana diffusion, we do not have an estimate.
The table also includes orbit circulation time, $\tau_o$, estimated analytically, and passing particle loss time, $\tau_s$, estimated from the numerically computed advection-diffusion coefficients.

\begin{table}[t]
\centering
\resizebox{0.49\textwidth}{!}{
\begin{threeparttable}
\caption{Relevant timescales for trapped particle transport.}
\label{tab: timescales}
\begin{tabular}{>{\kern-\tabcolsep}*{3}{m{1.1cm} p{2.6cm} p{2.8cm}}<{\kern-\tabcolsep}}

\toprule
Symbol   & Process                              & Estimate \\ \midrule
\multicolumn{1}{c}{$\tau_c$} & Pitch scattering & 
$\nu^{-1}$ from Eq.~\eqref{eq: pitch scattering} \\
\rowcolor[HTML]{EFEFEF}
& & \\
\rowcolor[HTML]{EFEFEF}
\multicolumn{1}{c}{\multirow{-2}{*}{$\tau_e$}} & 
\multirow{-2}{=}{De-trapping due to electric field}   & \multirow{-2}{=}{$\frac{mc}{qE}p$} \\
& & \\
\multicolumn{1}{c}{\multirow{-2}{*}{$\tau_w$}} & 
\multirow{-2}{=}{De-trapping due to Ware pinch}   & \multirow{-2}{=}{$\frac{B_\theta}{E_\phi} R \left(\frac{r}{R_0} - \frac{\xi^2}{2-\xi^2} \right)^*$} \\
\rowcolor[HTML]{EFEFEF}
& & \\
\rowcolor[HTML]{EFEFEF}
\multicolumn{1}{c}{\multirow{-2}{*}{$\tau_o$}} & 
\multirow{-2}{=}{Orbit circulation time}  & 
\multirow{-2}{=}{$\frac{4\pi R q_s}{v \sqrt{2 r/R_0}}^{**}$} \\
\multicolumn{1}{c}{$\tau_s$} & Loss time & Eq.~\eqref{eq: loss time} \\
\bottomrule

\end{tabular}
\begin{tablenotes}
\item $^*$ $R_0$ is the major radius, $B_\theta$ poloidal field, and $E_\phi$ toroidal electric field.
\item $^{**}$ $v$ is particle velocity and $q_s$ is the safety factor.
\end{tablenotes}
\end{threeparttable}
}
\end{table}

\subsection{Trapped particle confinement}

A scan on RE initial parameters, pitch, energy, and radial position, was performed to assess the confinement of poloidally trapped REs.
Markers sharing the same initial pitch and energy values were distributed uniformly in radius at $t=12.5$~ms, i.e., right after the beginning of the stochastic phase, and traced for 1~ms.
If marker energy in the simulation was reduced below two times the local thermal energy, the marker was labelled thermalized and its simulation was ceased.
The simulation then was repeated with different values for the initial pitch and energy.

Four scans were done with different physics included and the results are shown in Fig.~\ref{fig: endstate}.
The case (a) with complete physics included electric field, full collision operator (i.e., both pitch and energy components were included), and radiation reaction force.
Scans with crippled physics were carried out to identify mechanisms causing trapped RE losses.
In (b), the pitch collision operator was disabled.
In (c), the collision operator was completely disabled.
In (d), the collision operator as well as the radiation reaction force were disabled and, additionally, we set $E=0$ everywhere.

For the analysis we have separated the momentum space into four regions, that are identified with a different symbol at the bottom-right corner of each panel.
In the \emph{energetic banana} region ($\bullet$) in the top-left corner we would not expect to see many particles as no REs are generated there via hot-tail mechanism, and it is unlikely that knock-on collisions would yield such energetic electrons.
Therefore, the only REs in this regime would be those that have scattered there from passing trajectories via the combined effect of pitch scattering and synchrotron losses.

The REs in this region are dominantly confined with the losses increasing with pitch and when the particle origin is closer to the core.
This observation along with the fact that the losses disappear when the electric field is switched off in (d), point out that the losses are mainly due to the Ware pinch.
The exception is the panel ($\xi=0.5$, $p = 100$) where orbits are widest, which makes these particles susceptible to the banana diffusion and losses are present even when the electric field is not.
The radiation reaction force does not seem to have an impact on transport.

In the final column we have \emph{passing REs} ($\blacktriangle$) that are all lost except for those that are initialized inside the axis region that has lower transport.
Near the critical momentum ($p_\mathrm{crit}\approx 0.1$) collisions also cool some of the passing particles, though this is more due to pitch scattering to trapped orbits than the collisional drag alone.

In the bottom row near the critical momentum we have the \emph{thermal region} ($\blacksquare$) where all electrons are thermalized.
Interestingly the picture changes completely if pitch collisions are disabled; in (b) and (c) all electrons are lost except for the ones near the axis.
All trapped electrons become confined again when electric field is disabled in (d), hence the electric field is strong enough to accelerate electrons to passing orbits during the time it takes to complete half a banana orbit, but pitch scattering interrupts this process.

In the \emph{trapped RE region} ($\bigstar$) we can expect to see REs either due to knock-on collisions or hot-tail generation.
All electrons in this region are either lost or thermalized, again with the exception of those located near the axis.
This is a sharp contrast to the energetic banana region where almost all REs were confined and warrants further study.
Losses are not present when the electric field is disabled in (d) but appear to some extent in the collisionless case (c) and are enhanced when collisions are included.
This indicates that losses are due to combined effect of pitch scattering and Ware pinch.
A curious feature in the middle row of (b) is the particles that are thermalized even though there are no thermalized electrons at lower energy.
These are particles that end up in passing orbits that are opposite to the electric field acceleration and remain there since there is no pitch scattering.
Only particles with high enough $p_\perp$ can become trapped again during the deacceleration. 
Otherwise they are thermalized by the combined effect of collisional drag and electric field acceleration that can happen before particles are lost by the stochastic field transport.

\begin{figure}[t]
\centering
\begin{overpic}[width=0.49\textwidth]{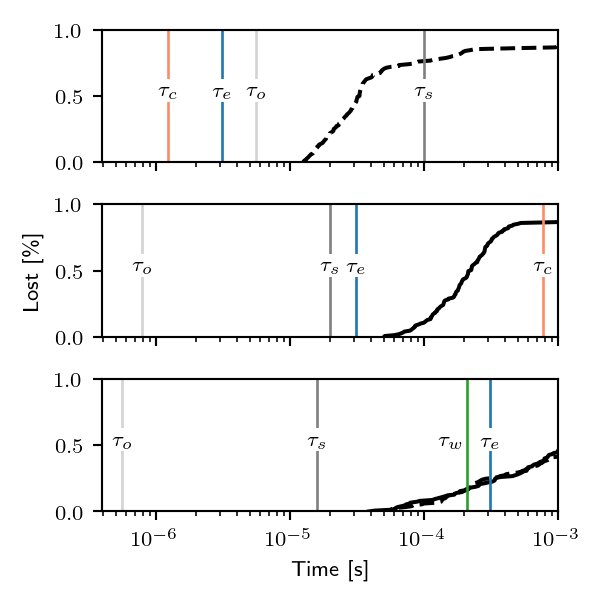}
\put(18,97){a) $p = 0.1$}
\put(18,68){b) $p = 1$}
\put(18,39){c) $p = 10$}
\end{overpic}
\caption{
Illustration of the time scales relevant for electron transport.
Each plot shows the cumulative losses for different electron initial energy as a function of time: with complete physics (solid black line) and with the collision operator disabled (dashed black line).
Annotated vertical lines show the characteristic times in each case.
For each case the initial pitch was $p_\parallel/p = 0.3$.
}
\label{fig: timescales}
\end{figure}

To provide support for our conclusions, Figure~\ref{fig: timescales} compares the loss rate in specific cases to the characteristic times we collected in Table~\ref{tab: timescales}.

The case (a) is for ($\xi=0.3$, $p = 0.1$), where particles were only lost when the electric field was included but collisions were disabled.
Here we see that the time it takes for the electric field to accelerate particles to the passing regime is shorter than the orbit time, making this mechanism possible.
However, the pitch scattering time is even shorter and, as such, collisions are able to disrupt the acceleration.
In the collisionless case, the losses do not appear immediately but roughly around the time scale for the stochastic field transport.

The case (b) shows the collisional scattering mechanism observed in panel ($\xi=0.3$, $p = 1$).
Now that the electric field acceleration time is larger than the orbit time, particles are not able to accelerate to passing orbits.
Instead, losses appear in the time scale corresponding to pitch scattering, which becomes even smaller than shown here as particles cool down.
Since the loss time is smaller than the pitch scattering time, passing particles are lost before they scatter back to the trapped regime.

The case (c) demonstrates the Ware pinch in panel ($\xi=0.3$, $p = 10$).
The pitch scattering time is now significantly longer than in (b), making the Ware pinch time the effective time scale.
Due to this, the same behavior is seen in both cases with and without collisions.
The fact that only 20~\% of the markers are lost till $t=\tau_w$ can be explained by noting that the estimate for $\tau_w$ assumes that the de-trapping occurs only once, when in Fig.~\ref{fig: orbits} it was seen that the particle goes over several cycles of trapping and de-trapping when the field is stochastic.
Therefore one should be mindful of using this estimate.

The orbit-following simulations performed in this work were done using the guiding center approximation.
We found that the main results were reproduced also when the whole gyro-orbit was solved.
The only difference was that losses due to trapped banana diffusion were somewhat higher in the guiding center picture.

\section{Summary and conclusions}
\label{sec:conclusions}

For the case studied here, the transport due to magnetic field stochasticity was sufficient to deconfine runaway electrons during early phase of current quench.
The stochastic phase begins after the beginning of the current quench and lasts for 8~ms, during which most of the plasma volume exhibits sufficient transport to mitigate the RE avalanche.
Close to the magnetic axis there is a possibility of a small fraction of the REs surviving.
However, the simulated case should not be taken as representative of all ITER current quenches, and the level of magnetic field stochasticity observed here might not hold in general.
For example, it is unknown what determines the duration of the stochastic phase.
Further work is required to assess how the stochasticity during the current quench depends e.g. on the initial conditions assumed for the MHD simulation.

Even though poloidally trapped REs are not directly affected by the field stochasticity, we identified three mechanisms that caused them to become deconfined as well.
These mechanisms and the energy range where they were dominant are: collisional scattering ($E_\mathrm{kin} \lesssim$ 200~keV), Ware pinch effect (200~keV $\lesssim E_\mathrm{kin} \lesssim$ 10~MeV), and collisionless banana diffusion ($E_\mathrm{kin} \gtrsim$ 10~MeV).
However, collisional scattering and Ware pinch do not cause losses directly  since these only push particles to the passing regime where they become promptly lost if the field is stochastic.
The confinement of trapped REs therefore depends not only on RE energy, electric field magnitude, and collisionality, but also on the magnetic field perturbation strength and the duration over which the field is stochastic.

For this work, the particle tracer in JOREK was retrofitted with operators for Coulomb collisions and radiation reaction force.
Collisional scattering was found to have significant impact on trapped particle dynamics and transport whereas the radiation reaction force did not.

This work does not completely address whether REs are mitigated in the studied ITER plasma.
One of the main unknowns is the distribution of REs generated during the thermal quench.
Further work is required in terms of kinetic modelling and MHD modelling of the thermal quench to overcome this issue, and later kinetic REs are required to be coupled to the MHD equations for an accurate assessment of beam evolution.

\ack
The authors are grateful to E. Nardon for providing valuable comments that improved the manuscript
This work has been carried out within the framework of the EUROfusion Consortium and has received funding from the Euratom research and training program 2014-2018 and 2019-2020 under grant agreement No 633053. The views and opinions expressed herein do not necessarily reflect those of the European Commission. Some of the simulations were done on the Marconi-Fusion supercomputer hosted at CINECA. ITER is the Nuclear Facility INB no. 174. This paper explores physics processes during the plasma operation of the tokamak when disruptions take place; nevertheless the nuclear operator is not constrained by the results presented here. The views and opinions expressed herein do not necessarily reflect those of the ITER Organization.
\appendix


\section*{References}
\bibliographystyle{iopart-num}
\bibliography{main}

\end{document}